\documentclass[12pt,a4paper]{iopart}

\usepackage{epsfig,times}

\def\msun{\mathrm{M}_\odot}
\def\Hz{\mathrm{Hz}}
\def\sec{\mathrm{sec}}
\def\snr{\mathrm{SNR}}
\def\fcut{f_{\mathrm{cutoff}}}
\newcommand{\htilde}{\tilde{h}}

\newcommand{\ambgty}[2]{\mathcal{H}(\Delta #1, \Delta #2)}

\begin{document}

\title[EHS algorithm for detection of gravitational waves]
      {Extended  hierarchical search (EHS) algorithm for detection of 
       gravitational waves from inspiraling compact binaries}
\author[*]{Anand S. Sengupta$^a$, Sanjeev V. Dhurandhar$^a$, 
           Albert Lazzarini$^b$ and Tom Prince$^c$}
\address{$^a$ Inter-University Centre for Astronomy and Astrophysics, 
Post Bag 4, Ganeshkhind, Pune 411 007, INDIA.}
\address{$^b$ LIGO Laboratory, California Institute of Technology, 
Pasadena CA 91125, USA.}
\address{$^c$ Jet Propulsion Laboratory, California Institute of Technology, 
Pasadena CA 91109, USA.}
\ead{anandss@iucaa.ernet.in, sanjeev@iucaa.ernet.in, lazz@ligo.caltech.edu, 
prince@caltech.edu}

\begin{abstract}

Pattern matching techniques like matched filtering will be used for online 
extraction of gravitational wave signals buried inside detector noise. This 
involves cross correlating the detector output with hundreds of thousands of 
templates spanning a multi-dimensional parameter space, which is very 
expensive computationally. A faster implementation algorithm was devised by 
Mohanty and Dhurandhar [1996] using a hierarchy of templates over the mass 
parameters, which speeded up the procedure by about 25 to 30 times. 
We show that a further reduction in computational cost is possible if 
we extend the hierarchy paradigm to an extra parameter, namely, the time of 
arrival of the signal. In the first stage, the chirp waveform is cut-off at a 
relatively low frequency allowing the data to be coarsely sampled leading to 
cost saving in performing the  FFTs. This is possible because  most of the 
signal power is at low frequencies, and therefore  the advantage due to 
hierarchy over masses is not compromised. Results are obtained for spin-less 
templates up to the second post-Newtonian (2PN) order for a single detector 
with LIGO I noise power spectral density. We estimate that the gain in 
computational cost over a flat search is about 100.
\end{abstract}

\submitto{\CQG}
\maketitle

\section{Introduction \label{section_intro}}

Massive compact binary  systems consisting of neutron stars (up to a distance 
of 20 Mpc) or black-holes (up to 200 Mpc) ranging in masses from 
$\sim$ 0.5 $\msun$ to $\sim$ 10.0 $\msun$ are one of the most promising 
sources for large scale laser interferometric detectors for gravitational 
waves (GW). Few minutes before the final merger, the GW frequency of such 
sources will lie within the bandwidth of the detectors. The expected GW 
waveform of these sources is known adequately to allow pattern matching 
techniques like matched filtering to be used for signal extraction. 

However matched filtering is computationally quite expensive, since hundreds 
of thousands of templates must be used to search the multidimensional 
parameter space. The hierarchical search algorithm of Mohanty and Dhurandhar 
\cite{monty1} which employed a hierarchy of templates over the mass parameters 
achieved a significant improvement in cutting down the cost by a factor 
between 25 and 30. 
In this paper we explore the possibility of extending the hierarchical 
paradigm to an extra dimension - the time of arrival of the signal. This is 
achieved by using the idea of decimating the signal in time (proposed before 
in \cite{tanaka} but not worked out in detail). This reduces the computational 
cost by about a factor of 100 over the 1-step search described in Section 
\ref{section_onestep} for a single detector with LIGO I noise power spectral 
density. 

\section{The 1-step search \label{section_onestep}}

In the stationary phase approximation, the Fourier transform $\htilde(f)$ 
of the  spin-less restricted second order post-Newtonian (2PN) 
waveform up to a constant factor $\mathcal{N}$, is given by
\begin{equation}
\htilde(f) = \mathcal{N} f^{-7/6} \exp i \left [  
                           -\frac{\pi}{4} - \Phi_0 + \Psi(f;M,\eta) \right ],
\end{equation} 
where $M$ is the total mass of the binary system, $\eta$ is the ratio of the 
reduced mass to the total mass, $\Phi_0$ is the initial phase at some fiducial 
frequency $f_a$, and $f$ is the frequency. $\mathcal{N}$ depends on 
\cite{onestep2} the masses, the distance to the binaries and $f_a$. 

The function $\Psi(f;M,\eta)$ describes the phase evolution of the 
inspiral waveform and is given by,
\begin{eqnarray}
&&\Psi(f;M,\eta) = 2\pi ft_0 + \frac{3}{128 \eta}
           \left [
            (\pi M f)^{-5/3} + \left ( \frac{3715}{756} + \frac{55}{9} \right )
            (\pi M f)^{-1}
            \right . \nonumber \\
           && \left . - 16\pi(\pi M f)^{-2/3}  
               + \left ( \frac{15293365}{508032} + \frac{27145}{504} \eta
              + \frac{3085}{72} \eta^2 \right )(\pi M f)^{-1/3} \right ]. 
\end{eqnarray}

\begin{figure}[t]
\centering
\includegraphics[width=0.45\textwidth]{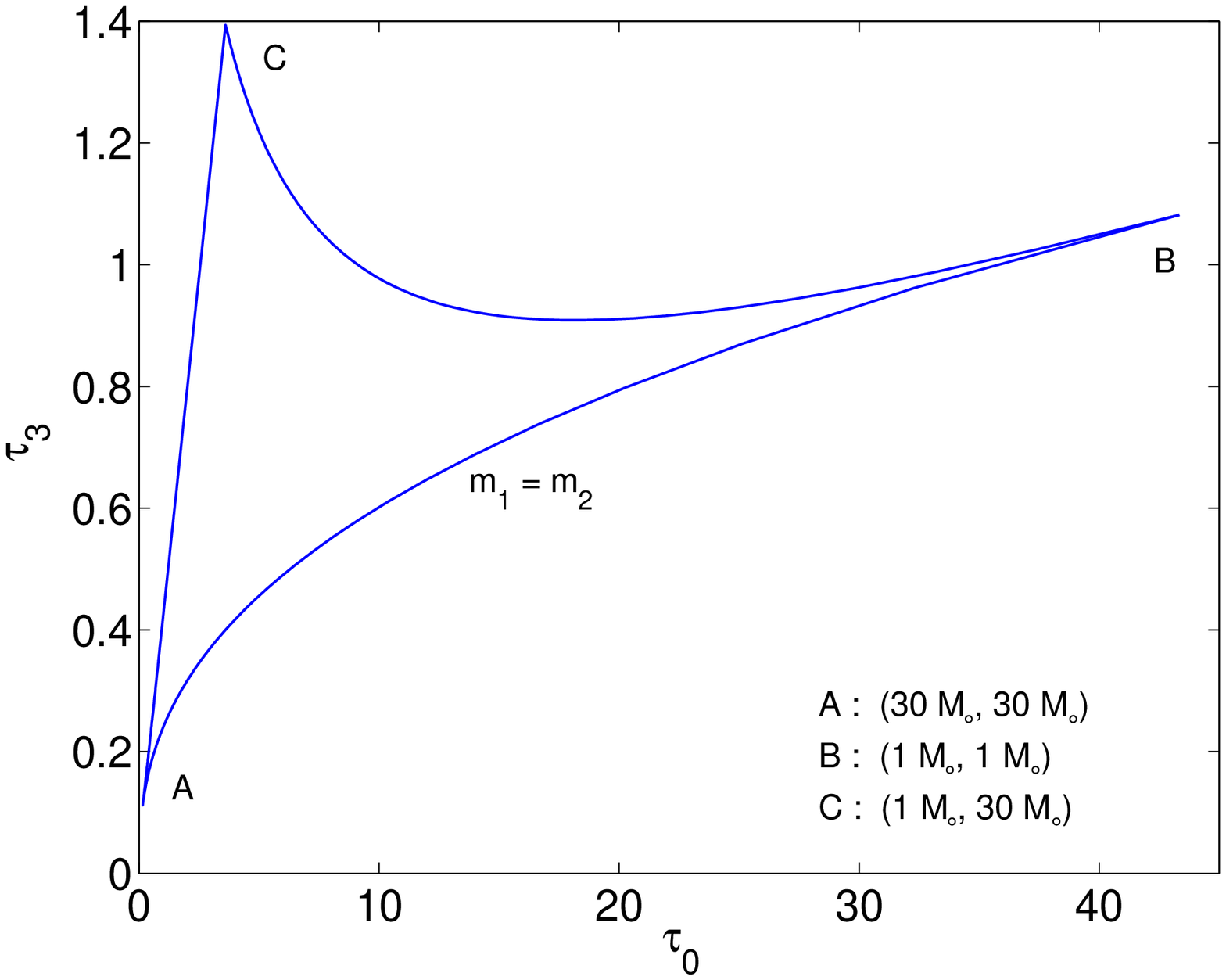}
\includegraphics[width=0.45\textwidth]{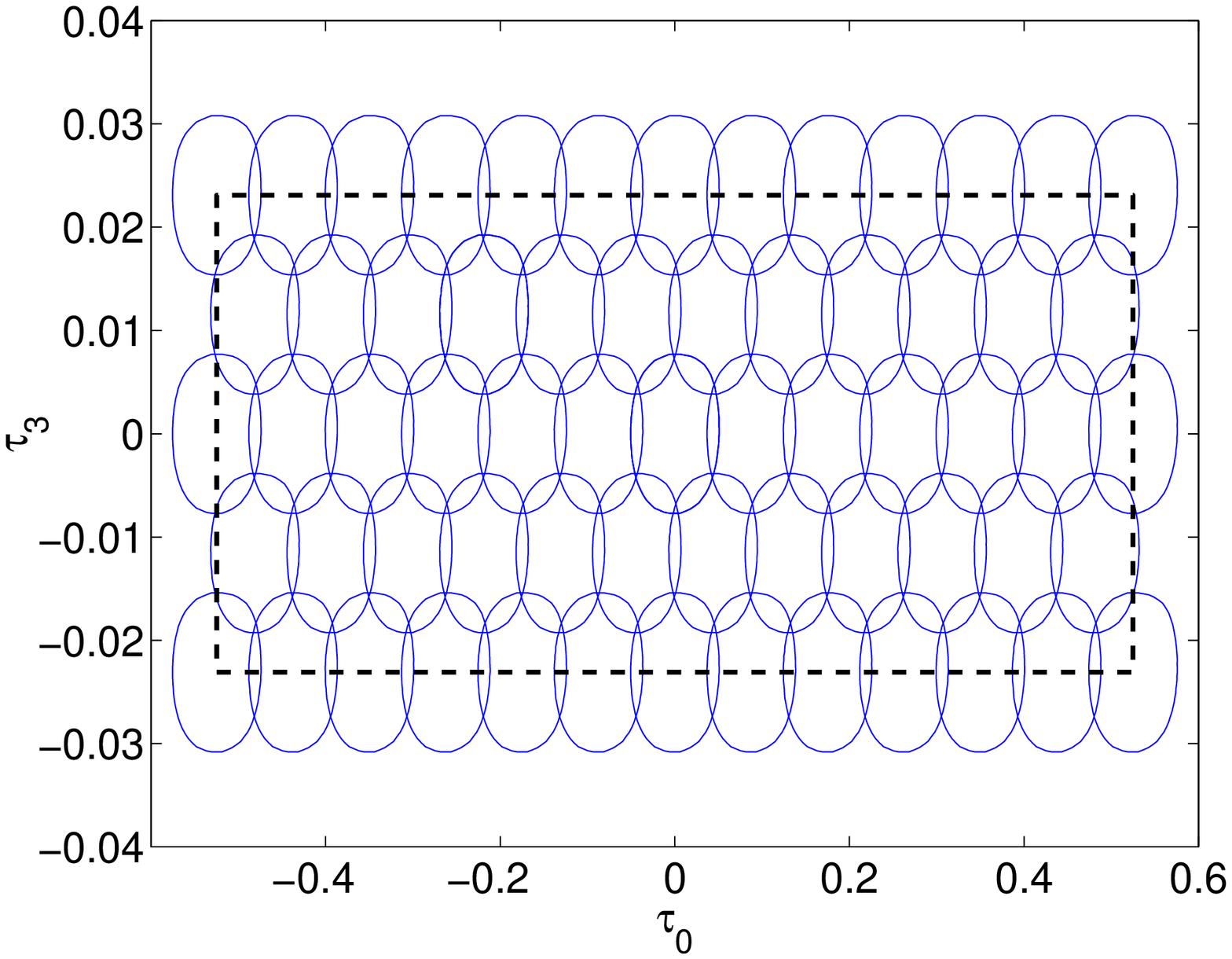}
\caption{Left panel shows the parameter space for 2PN waveforms in 
$(\tau_0,\tau_3)$ space for binary masses in the range 
$1.0\, \msun \leq m_1, m_2 \leq 30.0\, \msun$. 
The fiducial frequency is set to $f_a = 40.0$ Hz for this figure and the area 
is $8.5 \, \sec^2$.
Shown on the right panel is a schematic of discrete template placement using 
hexagonal closed packing ({\em hcp}). The spacing between neighboring 
templates is very small compared to the total area, and thus we can extend 
the method to the whole parameter space.}
\label{figarea}
\end{figure}

The kinematical parameters $(t_0, \Phi_0)$ do not determine the shape of the 
waveform and can be treated very simply  in the matched filtering procedure 
\cite{onestep1}. In what follows, the parameter space will explicitly refer to 
the space of dynamical parameters that determine the phase, and thus the shape 
of the waveform . For the 2PN case, it is the two dimensional space 
described by $M \, \mathrm{and} \, \eta$. In order to facilitate the analysis 
one chooses a new set of parameters  $\tau_0$ and $\tau_3$ in which  
the metric defined over the parameter  space \cite{owen} is approximately 
constant. They are related to $M$ and $\eta$ by
\begin{equation}
\tau_0 = \frac{5}{256\,\eta f_a} \left ( \pi M f_a\right )^{-5/3}, \,\,\,
\tau_3 = \frac{1}{8 \eta f_a} \left ( \pi M f_a \right )^{-2/3}.
\end{equation}
In these new co-ordinates the parameter space is wedge shaped as shown in the 
left panel of Figure \ref{figarea}.

A straight forward implementation of matched filtering involves constructing 
a discrete lattice of normalized templates over the parameter space and 
computing  correlations with the detector output. The bandwidth over which 
the correlations are computed in the frequency space, is taken from the 
seismic lower cutoff frequency $f_s$ to a upper cutoff frequency $f_c$. For 
the LIGO laboratory accepted noise PSD \cite{ligo_psd}, we set 
$f_s = 30\,\mathrm{Hz}$ and $f_u = 800\,\mathrm{Hz}$. The maximum of the 
correlation output is compared with the threshold $\eta_2$ (we put the 
subscript `2' because $\eta_2$ will be used as the second stage threshold
in the hierarchical scheme described later) which depends on a pre-decided 
false alarm rate usually set at one per year. The false alarm rate we assume 
here is for one detector only. If the data is sampled at  
$2048 \, \mathrm{Hz}$, the noise assumed to be a Gaussian stationary process 
and the number of templates $\sim 10^4$ then the threshold $\eta_2 \simeq 8.2$.
Detection is announced if the maximum correlation exceeds this threshold.

The degree of discretization of templates is governed by a maximum mismatch 
between the signal and the filter, which is taken to be $3\%$ (corresponding 
to a maximum loss of event rate of $10\%$).
The mismatch in terms of the parameters $(\tau_0, \tau_3)$ is encoded in the 
ambiguity function $\ambgty{\tau_0}{\tau_3}$. Actually $\mathcal{H}$ also 
depends weakly on the position $(\tau_0,\tau_3)$, but we ignore its effect in 
the present analysis. We use the most efficient hexagonal closed packing 
({\em hcp}) tiling for placing the templates as shown in Figure \ref{figarea} 
(right panel), with a packing fraction of $83 \%$. This leads to 17000 
templates for LIGO I noise PSD and for a mass range of 
$1.0 \, - \, 30.0 \, \msun$.  Sampling at $2048 \, \mathrm{Hz}$ with a data 
train $512$ sec long to allow for $75 \%$ padding, the number of points in 
the data train is $2^{20}$. The online speed required to cover the cost of 
FFTs amounts to 2.54 G-Flops using this 1-step implementation of matched 
filtering. The required speed  rises to about 15 G-Flops if the minimum mass 
of the mass range is reduced to $0.5 \msun$ and to about 150 G-Flop when the
minimum mass is $0.2\,\msun$.

\section{Extended hierarchical search}

The hierarchical scheme as given in \cite{monty1} involves parsing the data 
through two template banks instead of one.  The basic idea is to use a coarse 
bank of templates in the first stage along with a lower threshold. This allows 
the parameter space to be scanned with less number of templates.  However the 
lower threshold leads to a large number of false alarms which must be  then 
followed up with a fine bank (maximum mismatch $3\%$) search. 

\subsection{Methodology}

\begin{figure}[h]
\centering
\includegraphics[width=0.45\textwidth]{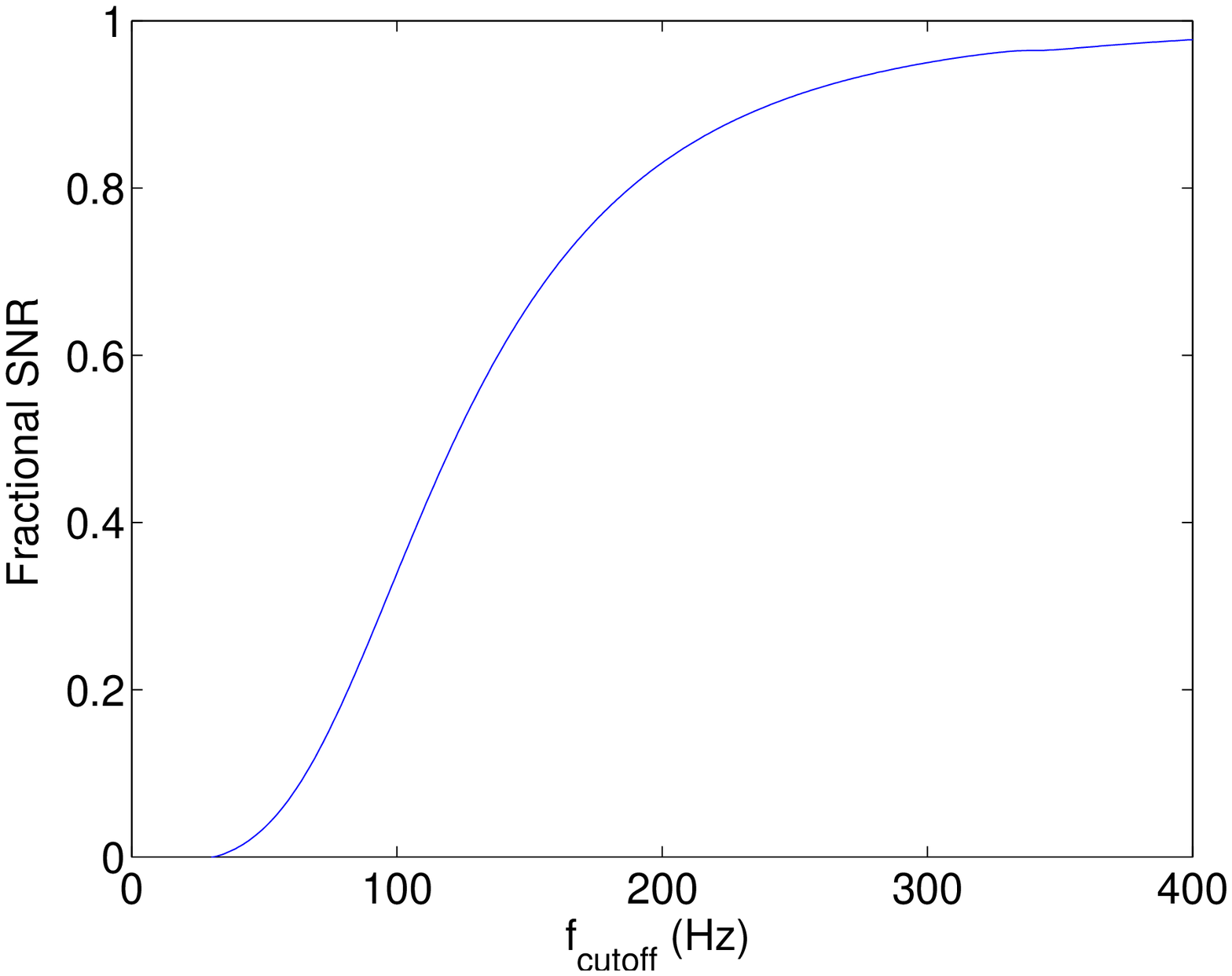}
\includegraphics[width=0.45\textwidth]{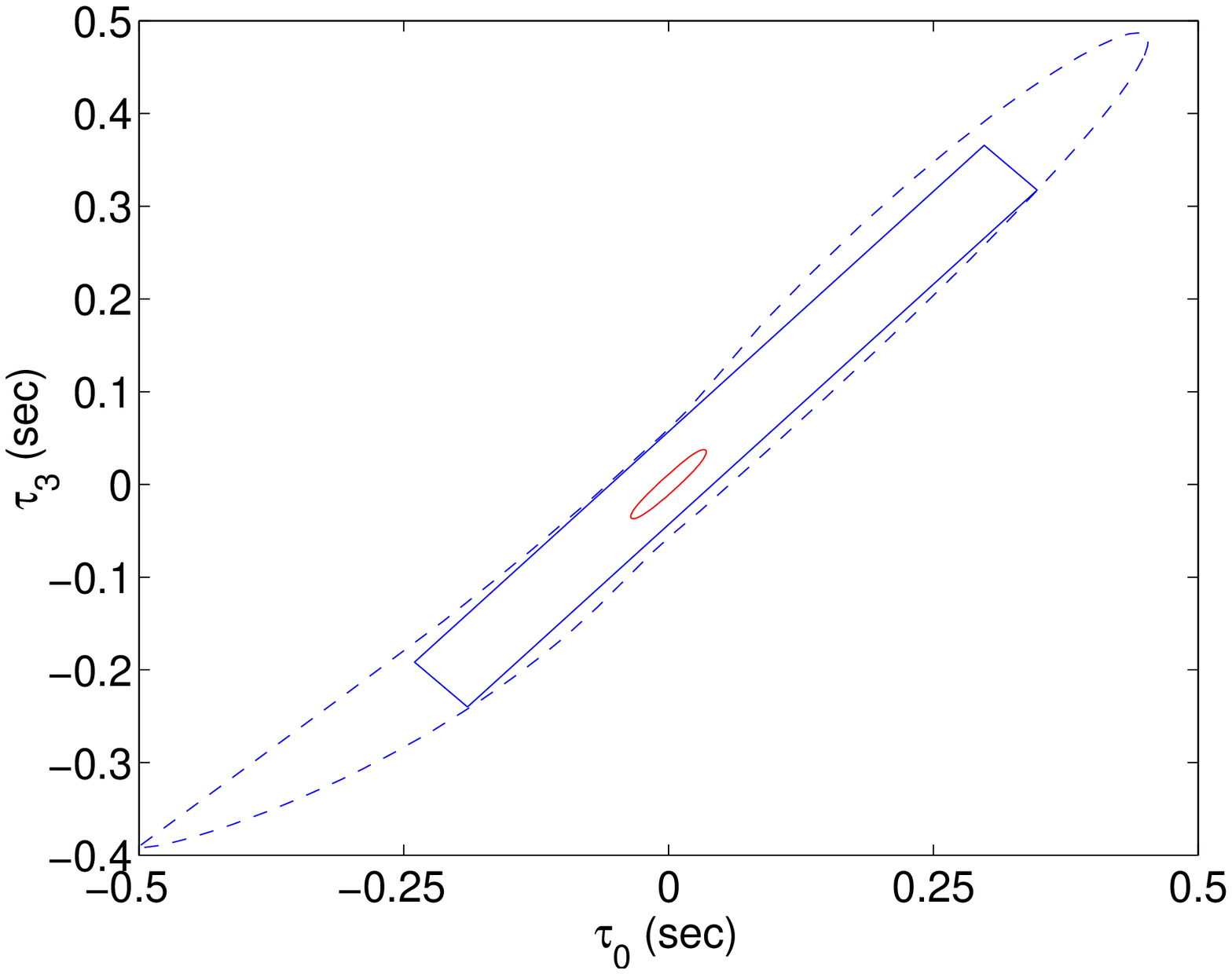}
\caption{The left panel shows the fraction of signal to noise ratio (SNR) 
extracted as a function of the cut off frequency $\fcut$. The SNR is 
normalized to 1 for $\fcut = f_u \,\Hz$. 
The right panel shows the $\ambgty{\tau_0}{\tau_3}= 0.80 $
contours along with the largest inscribed rectangle used in the trigger stage 
of EHS algorithm for $\fcut=256\,\Hz$. The ellipse inside the rectangle is 
that used by 1-step search to tile the space and is given for comparison of 
the areas.}
\label{figsnr}
\end{figure}

The above idea can be extended to include the time of arrival of the signal - 
by cutting off the chirp at a lower frequency $\fcut$ and re-sampling it at a 
lower Nyquist rate in the trigger phase of a two step implementation of 
matched filtering. This is possible because the inspiral signal contains 
most of the signal power at lower frequencies ( the power spectrum scales as
$\left | \htilde(f) \right |^2 \propto f^{-7/3}$) and thus we do not lose 
much of the SNR. The fractional SNR that can be recovered in this case is 
given by  $\snr(\fcut) = I(\fcut)\left /I(f_u) \right .$, where
\begin{equation}
I(\fcut) = 2 \int_{f_s}^{\fcut} \frac{df}{f^{7/3}\, S_h(f)} \,, \,\,\,\,\,
f_s < \fcut \leq f_u \, .
\end{equation}
In the left panel of Figure \ref{figsnr} we have plotted the relative SNR as a 
function of the cutoff frequency.
As can be clearly seen, a significant fraction of the signal power can be 
recovered from a relatively small value of $\fcut$. Specifically, for 
$\fcut = 256 \,\Hz$, almost $92 \%$ of the SNR can be recovered. Since not 
too much of the SNR is lost, the hierarchy over the masses \cite{monty2} is 
not  compromised too severely. 

The most obvious advantage of lowering the Nyquist rate in the first stage is 
that we have fewer points to contend with in computing the FFTs (the cost of 
FFTs scale as $N\log_2N$), leading to reduction in cost. For example, 
$\fcut = 256 \, \Hz$ reduces the sampling rate to a quarter, reducing  the 
cost of FFTs by almost the same factor. Secondly, the ambiguity function 
becomes wider thereby reducing the number of templates used in the first 
stage and reducing the cost. 

The first stage threshold $\eta_1$ is set by striking a balance between the 
two opposing effects; (i) $\eta_1$ must be low enough to allow coarsest 
possible placement of trigger stage templates so that the first stage 
computational cost is minimized,
(ii) $\eta_1$ must be high enough to reduce the number of false crossings,  
so that the second stage cost is reduced.

By choosing $\eta_1$ optimally, the total computational cost arising from 
both the stages is minimized.
 
\subsection{Computational cost of the EHS}

We consider data trains of length $512 \, \sec$ which accommodate chirps of 
maximum length $95\,\sec$ with sufficient padding corresponding to minimum 
mass limit of $1.0 \,\msun$. We consider first and second stage sampling 
frequencies of $512\,\Hz$ and $2048\,\Hz$ respectively. The cutoff frequency 
in the first stage is set at $\fcut = 256\,\Hz$. Thus the data points to be 
processed are $2^{18}$ and $2^{20}$ for the coarse and fine banks respectively.
The second stage threshold is set at $\eta_2 \simeq 8.2$ as before 
(see Section \ref{section_onestep}) for a false alarm rate of one per year. 
The signals of minimal strength that can be observed, assuming a minimum 
detection probability of $95\%$ and $3\%$ mismatch between templates turns 
out to be $9.17$. The first stage $\snr$ for the above mentioned cutoff can 
be calculated to be $92\%$ of 9.17 which is  8.44. It is found that the total 
cost is minimized for  $\eta_1 \approx 6$. Again, assuming a minimum detection 
probability of $95\%$ in the first stage itself, the ambiguity function can be 
allowed to drop to $6.75$ which is about $80\%$ of 
its maximum value. This large drop allows for very big tiles as shown in the 
right panel of Figure \ref{figsnr}. However, the awkward shape of these tiles 
makes tiling the parameter space a difficult proposition. We surmount this 
problem by cutting out the largest inscribed rectangle as shown in the same 
figure which has an area of $0.05\,\sec^2$. This is $\sim 50$ times larger 
than the area of second stage tiles.
Combining the first and second stage costs, the required online speed is then 
estimated at 19.54 M-Flops, giving a gain factor of about 130 over the one step
search.
  
\section{Discussion}
\begin{figure}[h]
\centering
\includegraphics[width=0.45\textwidth]{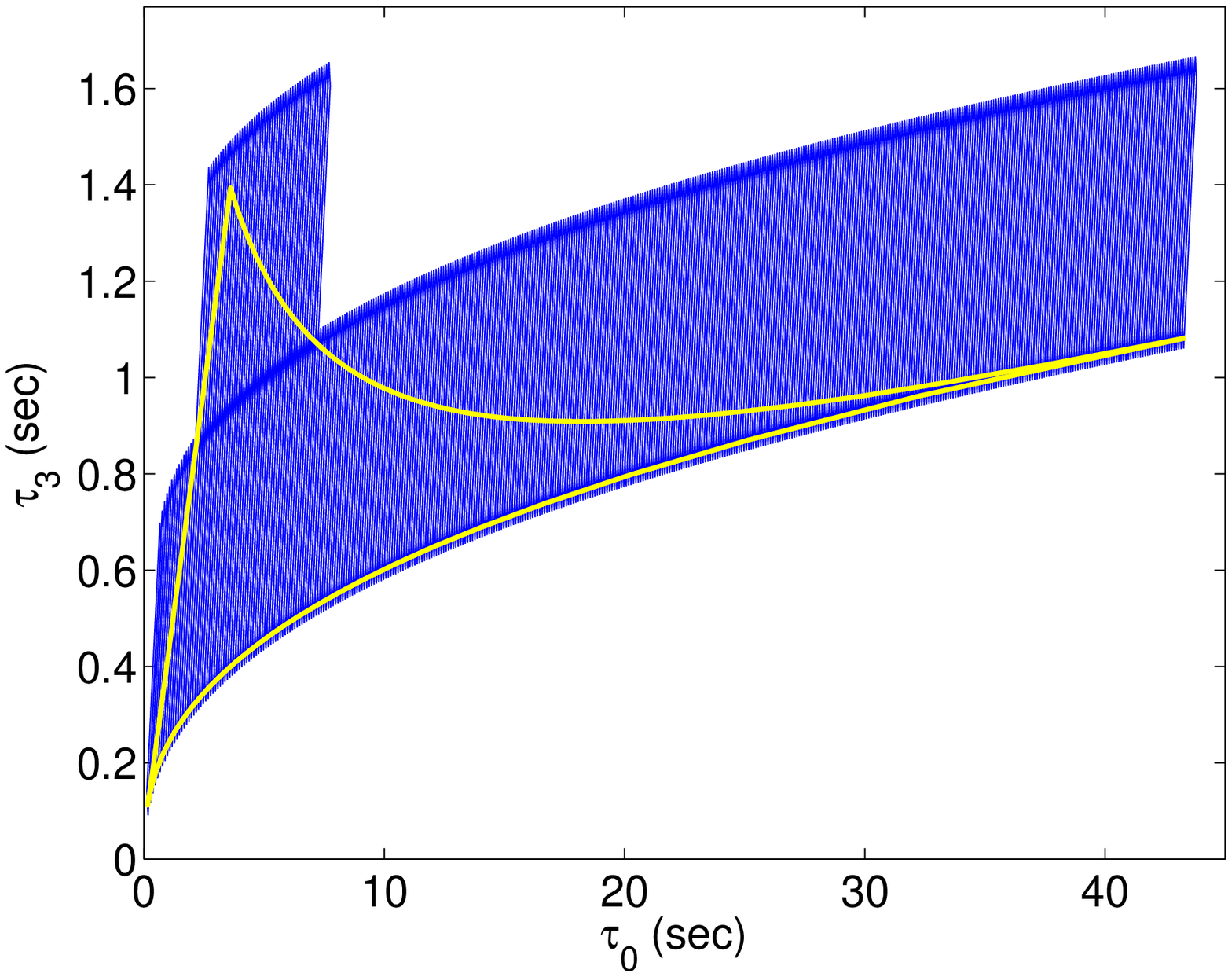}
\includegraphics[width=0.45\textwidth]{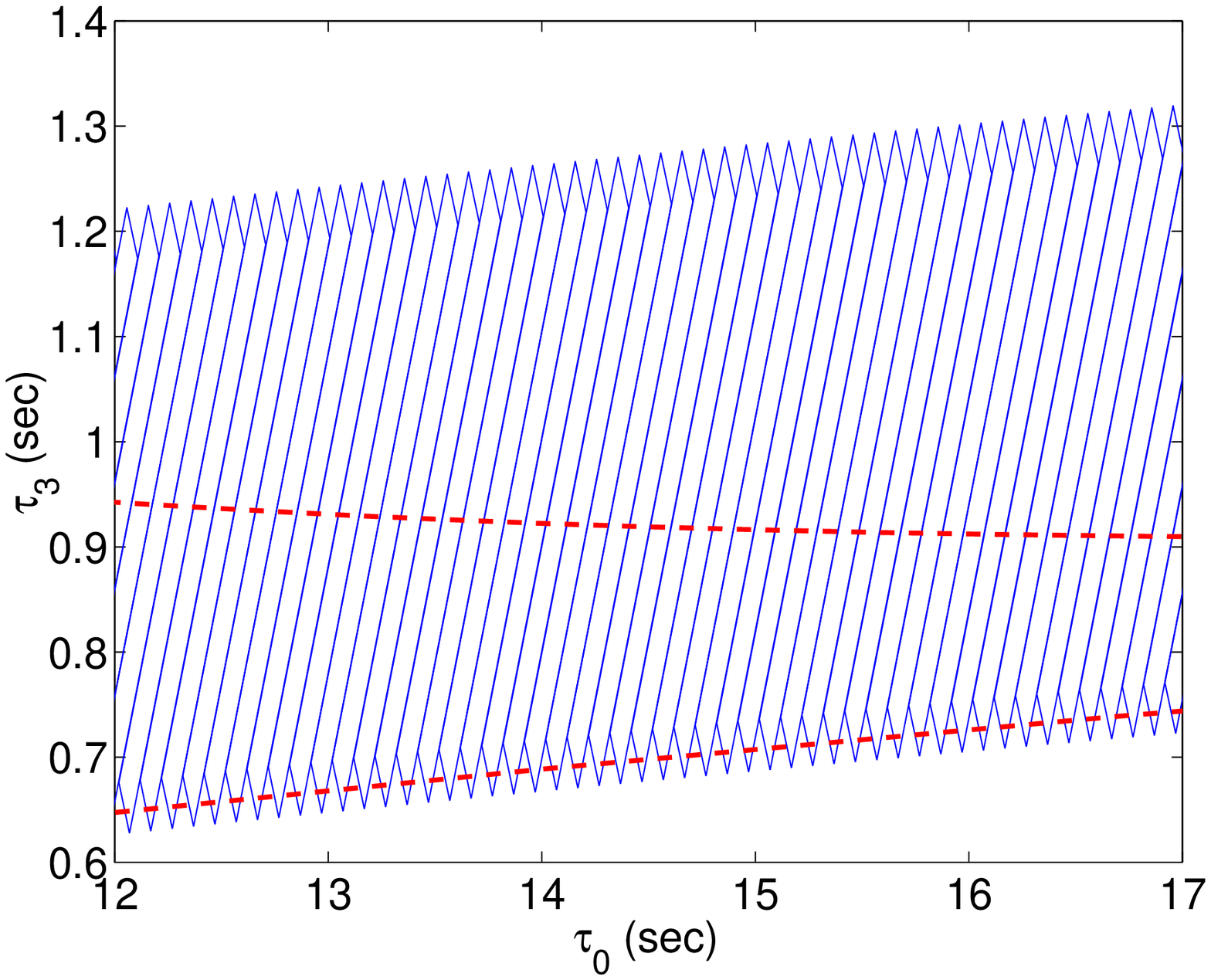}
\caption{The left panel shows a possible schematic lay-out of first stage 
templates for implementing the EHS algorithm.
Note that far too many templates have to be used in the first stage to cover 
the parameter space, due to inefficient tiling. The effect is most severe 
towards the tapering low mass end of the parameter space. 
The right panel shows the details of first stage template placement at a higher
resolution. The dashed line in this figure is the parameter space boundary.}
\label{figlay}
\end{figure}    

We have shown that a three dimensional hierarchical search for gravitational
waves emitted by inspiraling compact binaries is a 
highly promising proposition. 
We have obtained a computational cost reduction factor of about 130 for an 
ideal search, when the parameter space can be optimally tiled by 
the first stage template rectangles. However due to {\em boundary effects} 
this factor can be drastically reduced. As can be seen in Figure \ref{figlay}, 
the parameter space tapers towards the low mass end where the tiling becomes 
inefficient. 
Most of the tiles go `out' of the deemed parameter space. It is estimated that
this can cut down the above mentioned gain factor  by about $50\%$. These 
effects become less
pronounced when the lower mass limit of the mass range is 
reduced. We expect to recover the gain factor to about 100, if the lower
mass limit is reduced to less than $0.5\,\msun$.

The second important effect, affecting the gain factor, will come from the
rotation of the first and second stage tiles and also the differential rotation
between them.  These effects have still to be taken into account and will be
the thrust of our future investigations. However, we do not expect this effect 
to alter the gain factor very much.

\ack

The authors would like to thank A. Vicere, B. S. Sathyaprakash, S. Mohanty 
for fruitful discussions. S.V. Dhurandhar would like to thank B. Barish, G. 
Sanders for a visit to Caltech. A. Sengupta would like to thank CSIR for JRF. 
This work has been partially supported by LIGO Laboratory under NSF 
cooperative agreement PHY92-10038. This document has been assigned LIGO 
Laboratory document number LIGO-P010020-00-E ..

\end{document}